# Using Di quark pairs as a way to obtain dark matter near the beginning of nucleation of a new universe

*A. W. Beckwith*
*Department of Physics and*
*Texas Center for Superconductivity and Advanced Materials*
*University of Houston*
*Houston, Texas 77204-5005, USA*

**Abstract**

Using the Bogomil'nyi inequality and the vanishing of topological charge at the onset of nucleation of a new universe permits a simpler, more direct use of how topological defects (kinks and anti kinks) contribute to the onset of initial conditions at the beginning of inflationary cosmology. We find that it is possible to present inflationary cosmology with initial conditions with di-quark pairs formed using an axion wall as a way to start formation of di quarks as soliton – anti soliton type pairs(S-S') . This allows for a transition to inflationary cosmology congruent with an initial nucleation state of matter which would have a significant dark matter component in the beginning. The separation between the di quark pair constituents would be due to CP violations in the initial phases of matter creation and would be a convenient trigger for inflationary expansion.

**PAC numbers   02.30. Em**, **03.75.Lm, 11.27.+d, 98.65.Dx , 98.80.-k**

## I. Introduction

I wish to present a different paradigm as to how topological defects ( kinks and anti kinks) contribute to the onset of initial conditions at the beginning of inflationary cosmology. Currently as seen by two papers written by Mark Trodden[1] and Trodden et al[2]. topological defects are similar to D branes of string theory [1]. While this soliton (anti soliton) construction permits extensions to various super symmetric theories, it also obscures direct links to inflationary cosmological potentials such as Guth's harmonic potential [3].

The zeroth level assumption underlying this is that there could be a CP violation in the initial phases of states of matter. This in turn leads to Baryon matter state separation into Baryon-anti Baryon pairs (di quark pairs) which in turn would lead to the soliton-anti soliton pair formation alluded to in this particular paper. If the di quark pairs form we would then have a situation where an overall topological charge Q would tend to then vanish for our physical system.

Since I wish to make the linkage clearer, I present how soliton- anti soliton (**S-S'**) di quark pairs can be an initial starting point for times $t \leq t_P$ where $t_P$ is Plancks discretized smallest unit of time , as a coarse graining of time stepping in cosmological evolution. Initially I describe work which has been done by Zhitnitsky about formation of a soliton object via a so called di quark condensate[4]. This will lead to a scalar potential formation of soliton- anti soliton pair structures which help form a scalar field . The di quark pairs help form a scalar field $\phi$ whose dynamics are described by a driven Sine Gordon potential [5]which has much of its potential contribution eliminated in the times immediately after $t_P$, due to the vanishing of topological charge Q [5]. Afterwards, a 2nd potential regime is initiated which is proportional to the square of a scalar field $\phi$ divided by a denominator of the form $(1 + A \cdot \phi^3)$ [6,7] which in turn blends into the square of a scalar field $\phi$ for a 3rd potential regime corresponding to Guth's chaotic inflationary model [3]. I should stress that this 2nd potential regime is actually akin to what happens in typical inflationary models, with an initially large potential forming a potential barrier at the onset of time $t \approx t_P + \delta \cdot t$ and then decreasing into the third potential regime proportional $\phi^2$ for times $t >> t_P$ [3] and can be observed in figure 6.6 [8] ( scalar field trapped in a false vacuum) of Dodelsons book .

    This paper starts off with several premises (assumptions)

0th , that a CP violation in initial states would lead to an initial Baryon condensate of matter separating into di quark ( S-S') pairs leading to :

1st ,that for times less than or equal to Planck time $t_P$ the potential system for analyzing the nucleation of a universe is a driven Sine Gordon system [9], with the driving force in magnitude far less than the overall classical Sine Gordon potential.

2nd premise lies in having topological charges for a soliton – anti soliton di quark pair $(S - S')$ stem prior to Planck time $t_P$ for this potential system cancel out, leaving a potential proportional to $\phi^2$ minus a contribution due to quantum fluctuations of a scalar field being equal in magnitude to a classical system, with the remaining scalar potential field contributing to cosmic inflation in the history of the early universe.

The 3$^{rd}$ assumption is that a vacuum fluctuation of energy equivalent to $\Delta t \cdot \Delta E = \hbar$ will lead to the nucleation of a new universe, provided that we are setting our initial time $t_P \approx \Delta t$ as the smallest amount of time which can be ascertained in a quantum universe.

What has to be kept in mind is the following. If a phase transition occurs right after our nucleation of an initial state, it is due to the time of nucleation actually being less than (or equal to) Planck's minimum time interval $t_P$, with the length specified by reconciling the fate of the false vacuum potential used in nucleation with a Bogomol'nyi inequality specifying the vanishing of topological charge. I use di-quark (**S-S'**) pairs to represent an initial scalar field which after time $t_P \approx \Delta t$ will descend into the typical chaotic inflationary potential used for inflationary cosmology.

## II. Including in necessary and sufficient conditions for forming condensate state at or before Planck time $t_P$

We need to look at Ariel Zhitnitsky's formulation of how to form a condensate of a stable soliton style configuration of cold dark matter as a starting point for how an axion field can initiate forming a so called QCD ball[4]. This ball could in fact be the template for the initial expansion of a scalar field leading to false vacuum inflationary dynamics in the expansion of the universe. Since this is non standard material, I will outline the premises of the work being referred to. Zhitnitsky's formulation uses quarks in a non hadronic state of matter, but which in the beginning can be in di quark pairs. A di quark pair would permit making equivalence arguments to what is done with cooper pairs and a probabilistic representation as to find the relative 'size' of the cooper pair. We assume that something analogous can be done with respect to di quark pairs. In doing so, Zhitnitsky's calculations gave for quarks being squeezed by a so called QCD phase transition due to the violent collapse of an axion domain wall. The axion domain wall would be the squeezer to obtain a so called soliton (anti-soliton) configuration. This pre supposes a formation of a highly stable soliton type configuration in the onset due to the growth in baryon mass

$$M_B \approx B^{8/9}$$

(1)

This is due to a large baryon(quark) charge $B$ which Zhitnitsky says is smaller than an equivalent mass of a collection of free separated nucleons with the same charge[4]. And, Zhitnitsky further gives a criteria for absolute stability by writing a region of stability for the QCD balls dependent upon the inequality occurring for $B. > B_C$ (a critical charge value)

$$m_N > \frac{\partial M_B}{\partial B} \qquad (2)$$

He furthermore states that stability, albeit not absolute stability is still guaranteed for the formation of meta stable states occurring with

$$1 << B < B_C \qquad (3)$$

If one makes the assumptions that there is a balance between Fermi pressure $P_f$ and a pressure due to surface tension, with $\sigma$ being an axion wall tension value[4], so that

$$\left(P_\sigma \cong \frac{2\sigma}{R}\right) \equiv \left(P_f \cong -\frac{\Omega}{V}\right)$$

(4)

This pre supposes that $\Omega$ is some sort of thermodynamic potential of a non interacting Fermi gas, so that one can then get a mean radius for a QCD ball at the moment of formation of the value, when assuming $\tilde{c} \approx .7$, and also setting $B \approx B_C \propto 10^{+33}$ so that

$$R \equiv R_0 \cong \left(\frac{\tilde{c} \cdot B^{4/3}}{8 \cdot \pi \cdot \sigma}\right)^{1/3} \tag{5}$$

If we wish to have this of the order of magnitude of a Planck length $l_P$, then the axion domain wall tension must be huge, which is not unexpected. Still though, this pre supposes a minimum value of $B$ which Zhitnitsky set as

$$B_C^{\exp} \sim 10^{20} \tag{6}$$

We need to keep in mind that Zhitnitsky set this parameterization up to account for a dark matter candidate. I am arguing that much of this same concept is useful for setting up an initial condensate of di quark pairs as, separately soliton(anti-soliton) in the initial phases of nucleation, with the further assumption that there is an analogy with the so called color super conducting phase (CS) which would permit di quark channels, so that the problem we are analyzing not only is equivalent to BCS theory electron pairs, but can be linked to creating a region of nucleated space in the onset of inflation which has solution-anti soliton(S-S') pairs. The S-S' pairs would have a distance between them proportional to distance mentioned earlier, $R_0$, which would be greater than or equal to the minimum Plancks distance value of $l_P$ mentioned earlier. This will be useful when examining a starting point for the

$$\begin{array}{lll} V_1 & \rightarrow V_2 & \rightarrow V_3 \\ \phi(increase) \leq 2 \cdot \pi & \rightarrow \phi(decrease) \leq 2 \cdot \pi & \rightarrow \phi \approx \varepsilon^+ \\ t \leq t_P & \rightarrow t \geq t_P + \delta \cdot t & \rightarrow t >> t_P \end{array} \tag{7}$$

transition which will be mentioned later on. The potentials $V_1$, $V_2$, and $V_3$ will be described in terms of S-S' di quark pairs nucleating and then contributing to a chaotic inflationary scalar potential system.

## III Chaotic inflationary scenarios and their tie in with our problem of false vacuum nucleation.

Guth [3] as of 2000 wrote two well written articles with regards to the problem of the basic workings of inflationary models. The simplest of these models, called the chaotic inflationary model [3] via use of a massive scalar field construction gives an elegant treatment of how we could have an inflation field $\phi$ set at a high value $\phi \equiv \tilde{\phi}_0$ and which then would have an inequality of [3]

$$\tilde{\phi}_0 > \sqrt{\frac{60}{2 \cdot \pi}} M_P \approx 3.1 M_P \tag{8}$$

This pre supposes a harmonic style potential of the form [3]

$$V \equiv \frac{1}{2} \cdot m^2 \cdot \phi^2 \tag{9}$$

where we have classical and quantum fluctuations approximately giving the same value for a phase value of [3]

$$\phi^* \equiv \left(\frac{3}{16 \cdot \pi}\right)^{\frac{1}{4}} \cdot \frac{M_P^{3/2}}{m^{\frac{1}{2}}} \cdot M_P \to \left(\frac{3}{16 \cdot \pi}\right)^{\frac{1}{4}} \cdot \frac{M_P^{3/2}}{m^{\frac{1}{2}}} \tag{10}$$

where we have set $M_P$ as the typical Plancks mass which we normalized to being unity in this paper for the hybrid false vacuum – inflaton field cosmology example, as well as having set the general evolution of our scalar field as having the form of [3]

$$\phi \equiv \tilde{\phi}_0 - \frac{m}{\sqrt{12 \cdot \pi \cdot G}} \cdot t \qquad (11)$$

We assume that after an interval of time greater than Plancks unit of time that a nucleation of an initial universe evolve toward chaotic inflationary conditions.

## IV   Description of the potential used for nucleation and its blending into chaotic inflationary cosmology.

I look at reasonable potentials which would incorporate some of the insights of the chaotic inflation model ( $\phi^2$ potential dependence ) with false vacuum nucleation. For this potential, I worked with [10,11] :

$$V_1(\phi) = \frac{M_P^2}{2} \cdot (1 - \cos(\phi)) + \frac{m^2}{2} \cdot (\phi - \phi^*)^2 \qquad (12)$$

where $M_P > m$ as well as an overall potential of the form

$$V(\phi) \equiv \left[ initial \quad energy \quad density \right] + V_1(\phi) \qquad (13)$$

where what I am calling the initial energy density is a term from assuming a brane world type of potential usually written as [11]

$$V(\phi, \tilde{\psi}) = \frac{1}{4} \cdot \left( \tilde{\psi}^2 - M_P^2 \right)^2 + \frac{1}{2} \cdot \lambda' \cdot \phi^2 \cdot \tilde{\psi}^2 + V_1(\phi) \qquad (14)$$

where the 'radial' component $\tilde{\psi}$ is nearly set equal to zero and the scalar potential in our case is changed from a $\phi^2$ potential dependence to one where we incorporate a false vacuum nucleation procedure as given by $V_1(\phi)$. I will look at setting values of $\phi \equiv \phi^*$ [1] due to the chaotic inflation model [3] and then consider a specific ratio of $M_P$ to mass m to work with This same value of the inflaton field will lead to , as seen in **Figure 1b** , true

and false vacuum minimum values [12] when we are, here, using the Bogomil'nyi inequality [5, 13] with

$$V_1(\phi_F) - V_1(\phi_T) \cong .373 \propto L^{-1} \tag{15}$$

namely

$$\frac{(\{\ \})}{2} \equiv \Delta E_{GAP} \equiv V_1(\phi_F) - V_1(\phi_T)$$

and

$$\{\ \} \equiv \{\ \}_A - \{\ \}_B \equiv 2 \cdot \Delta E_{GAP} \tag{16}$$

where I will, for this cosmological example, set :

$$\{\ \}_A \approx M_P^2 + 2 \cdot m^2$$
$$\{\ \}_B \approx \frac{2 \cdot M_P^2 \cdot \phi_T \cdot \phi_F}{3 \cdot !} \tag{17a,b}$$

I am assuming that the net topological charge will vanish and that for a D+1 dimensional model with phenomenology equivalent to quasi one dimensional behavior due to near instantaneous nucleation that I work with the situation as outlined in **Figure 1a**, earlier where the quantity in brackets is set by **Figure 1b**. The details of that pop up are such that I am assuming a toy model with a thin wall approximation to a topological S-S' pair equivalent to assuming that the false vacuum paradigm of Sidney Coleman [12], ( as well as Lee and Weinbergs topological solitons associated with a vacuum manifold SO(3)/ U(1) [1, 2, 5, 14] ) holds in the main part.

The 2$^{nd}$ part of this potentials behavior scales as

$$V_2(\phi) \approx \frac{(1/2) \cdot m^2 \phi^2}{(1 + A \cdot \phi^3)} \tag{18}$$

which corresponds to a decreasing scalar field $\phi$ right after times $t \geq t_P + \delta \cdot t$ and the beginning of the collapse of the total charge Q of this kink – anti kink ensemble expanding into inflationary cosmological potential behavior. This then will lead to [3]

$$V_3(\phi) \approx (1/2) \cdot m^2 \phi^2 \tag{19}$$

with the bridge between these three regimes of the form given in equation 7 already.

## V. Conclusions

This article has laid out the case that the false vacuum hypothesis [12] gives a necessary condition for considering transport between adjacent but varying in magnitude local minimum values for the generalized potential split into the three parts indicated by this manuscript. . For the 1st potential the difference in the relative energy levels of the local minimum leads to, by the Bogomil'nyi inequality [5,12] to conditions of a nucleating S-S' pair . The 2nd potential corresponds to a decaying scalar field but initial potential hill corresponding to the topological charge vanishing in the onset of inflation, whereas the 3rd potential is the blend into traditional chaotic potentials as outlined by Guth[3] and others as a model for how chaotic inflation corresponds to approaching a global minimum as a ground state of inflationary cosmology which is observable. I assume that the typical Heisenberg uncertainty principle, with respect to uncertainty in time and energy values leads to an initial pop up of a scalar potential to be in the neighborhood of the minimum of figure 1b, whereas afterwards topological charge of a nucleating potential pair vanishes after this a prioriti collapse leading to a reduction of scalar potential values leading to the behavior as designated by the 2nd and 3rd potentials given in the main manuscript above.

I have managed to find an argument for a newly nucleated universe to have a finite but quite small diameter as well as reconcile the chaotic inflationary model of Guth[3] with a new fate of the false vacuum paradigm for nucleation at the initial stages of the big bang. Hubble time . I am assuming for now that the initial 'radius' of a nucleating universe is largely answered by [15] $l_P^2/\lambda_S^2 \approx \alpha_{GAUGE} \approx e^{\phi}$, with the scalar dilaton field give by the arguments presented , with a peak value of $\phi \approx 2\cdot\pi$ .This means that the $\lambda$ is close to being of the order of the Plancks length at the beginning of quantum nucleation as well as at the end of the inflationary period. This also has implications as to the relative size of the universe, implying a radial 'distance' of the order of Plancks length $l_P$ in the initial nucleation phase . This is somewhat different from traditional calculations of this topic, about the initial starting point for inflationary expansion[16].

Several open questions remain though. The first is that CP violations are the driving influence leading to a constituent separation of initial matter states into baryon-anti baryon states. What could lead to this initial CP violation ? This would need to be investigated fully with respect to the known astrophysical mechanisms of particle physics behavior during inflationary cosmology. Secondly, this matter can be a way to explain the production of dark matter in the initial phases of inflationary expansion. How the CP violation question is answered will allow us to eventually numerically compute the percentage of dark matter could and should exist in the initial phases of inflationary cosmology. Trodden in 1997 used gauged topological solitons to circumvent the need for just an electroweak phase transition [17] , for explaining baryon asymmetry but still, especially with respect to nucleation at the beginning, we need to take into account the result from the theory of primordial neucleosynthesis requiring

$$2\times 10^{-10} < \eta \equiv \frac{n_b - n_{\bar{b}}}{s} < 7\times 10^{-10} \qquad (20)$$

This has $n_b$ as the number density of baryons, $n_{\bar{b}}$ for anti baryons, and $s$ as entropy density. Finding a mechanism for CP violations leading to a distance between di quark pairs in the early universe would be a real challenge [18], and the formation of di quark pairs in itself would be a way of minimizing a topological net charge, Q , in the di quark pairs.

.

**Figure captions** :

**FIG 1a** Evolution from an initial state $\Psi_i[\phi]$ to a final state $\Psi_f[\phi]$ for a double-well potential (inset) in a 1-D model, showing a kink-antikink pair bounding the nucleated bubble of true vacuum. The shading illustrates quantum fluctuations about the initial and final optimum configurations of the field , while $\phi_0(x)$ represents an intermediate field configuration inside the tunnel barrier. The upper right hand side of this figure is how the fate of the false vacuum hypothesis gives a difference in energy between false and true potential vacuum values which we tie in with the results of the Bogomil'nyi inequality

.

**FIG 1b** : Evolution from an initial state $\Psi_i[\phi]$ to a final state $\Psi_f[\phi]$ for a tilted double-well potential in a quasi 1-D cosmological model for inflation, showing a kink-antikink pair bounding the nucleated bubble of true vacuum.

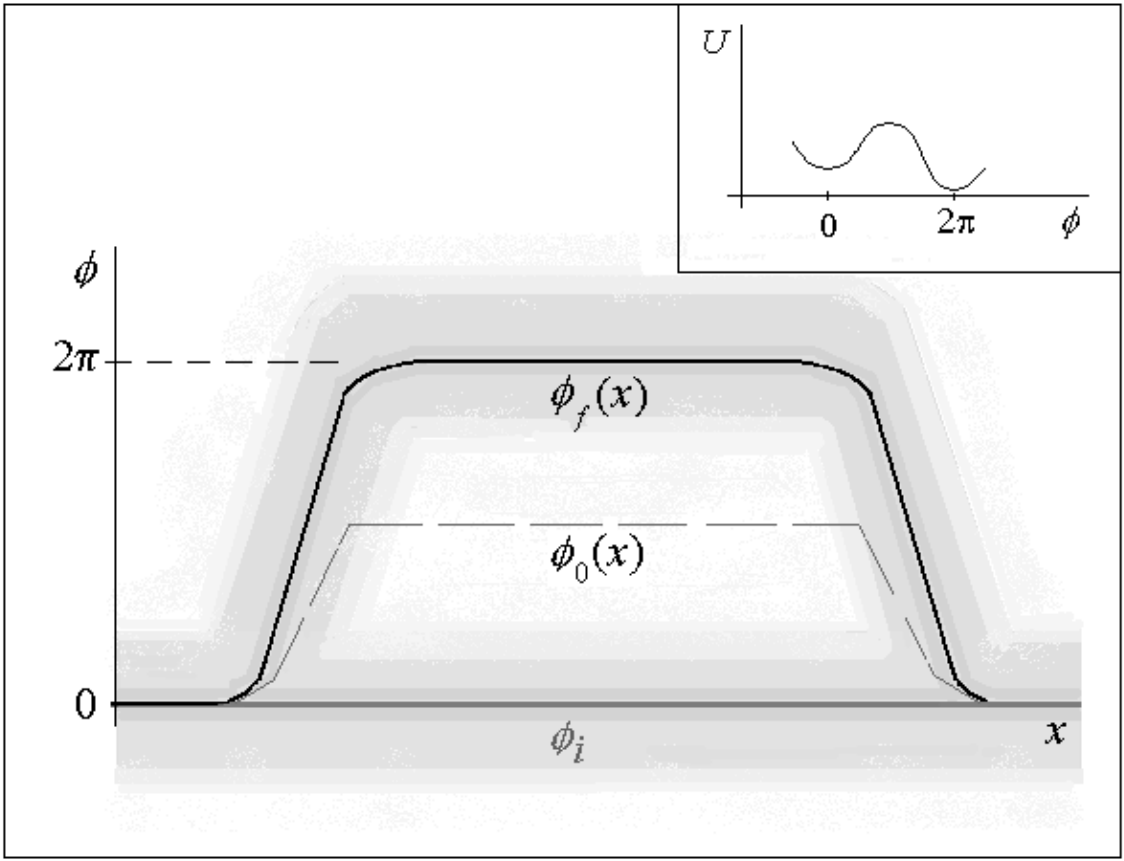

**Figure 1a**

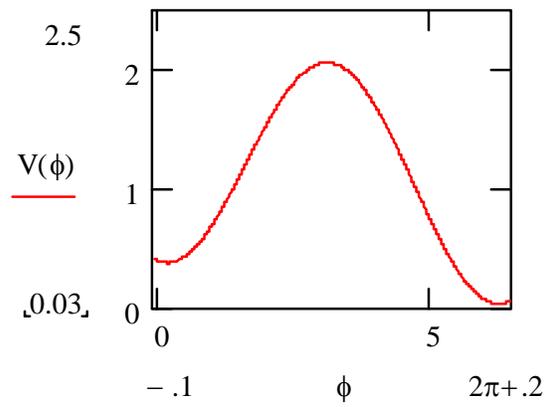

**Figure 1b**

**Beckwith**